# Realizing Wide Bandgap P-SiC-emitter Lateral Heterojunction Bipolar Transistors with low collector-emitter offset voltage and high current gain - A novel proposal using numerical simulation

M. Jagadesh Kumar[1] and C. Linga Reddy,
Department of Electrical Engineering,
Indian Institute of Technology, Delhi,
Hauz Khas, New Delhi – 110 016, INDIA.

Email: mamidala@ieee.org   FAX: 91-11-2658 1264

---

[1]Author for communication.

# Abstract

We report a novel method to reduce the collector-emitter offset-voltage of the wide bandgap SiC-P-emitter lateral HBTs using a dual-bandgap emitter. In our approach, the collector-emitter offset-voltage $V_{CE(offset)}$ is reduced drastically by eliminating the built-in potential difference between the emitter-base (EB) junction and collector-base (CB) junction by using a SiC-on-Si P-emitter. We demonstrate that the proposed dual-bandgap P-emitter HBT together with the SiGe base and Schottky collector, not only has a very low $V_{CE(offset)}$ but also exhibits high current gain, reduced Kirk effect, excellent transient response and high cutoff frequency. We evaluated the performance of the proposed device in detail using two dimensional device simulation and a possible BiCMOS compatible fabrication procedure is also suggested.

# 1. Introduction

Wide bandgap emitter bipolar transistors have several advantages over their homo-junction counterparts, such as (i) large current gain independent of emitter doping, (ii) increased base doping against current gain trade-off and (iii) high speed operation [1-3]. A wide bandgap emitter BJT can be realized either by reducing the base region bandgap with respect to the emitter or by increasing the emitter region bandgap with respect to the base. SiGe HBTs are an excellent example of wide bandgap emitter BJTs which have become very attractive in high speed applications due to their superior performance [4-5]. In the recent past, SiC as an emitter has also been shown to be a potential candidate for making wide bandgap emitter BJTs [6-8]. While other wide bandgap HBT technologies based on compound semiconductors such as InGaP/GaAs or AlGaAs/GaAs are available, HBTs based on SiC/Si are attractive because of their compatibility with the silicon technology and the excellent properties of SiC. In spite of a large lattice mismatch, wide bandgap SiC emitter hetero-bipolar transistors with large current gains have been successfully reported [6-8]. However, wide bandgap emitter HBTs exhibit a finite collector-emitter offset voltage, $V_{CE(offset)}$ [9-10] due to the large difference in the built-in potential of emitter-base and collector-base junctions. This is detrimental in many digital applications and should be minimized while retaining the advantages of the wide bandgap emitter.

A survey of literature reveals that only NPN wide bandgap SiC emitter transistors have been reported. Often, in many applications, compatible PNP transistors with wide bandgap SiC P-emitters are required. However, this is not possible because, as we shall demonstrate in the following section, SiC emitter PNP transistors are highly prone to the collector-emitter offset-voltage problems than the wide bandgap SiC emitter NPN transistors. If the $V_{CE(offset)}$ problem is eliminated in SiC wide bandgap emitter PNP transistors, they will find wide usage in many applications. To the best of our knowledge, no solution has yet been reported on how $V_{CE(offset)}$ can be minimized in wide bandgap emitter PNP transistors.

The aim of this paper is therefore to propose for the first time a novel method of minimizing the $V_{CE(offset)}$ in SiC wide bandgap emitter PNP transistors. We suggest that by making use of a dual-bandgap material consisting of SiC over Si in the emitter region will reduce the $V_{CE(offset)}$ without significantly affecting the current gain. We used numerical simulation to verify the efficacy of our solution to minimize $V_{CE(offset)}$. In our simulations, we have chosen a lateral experimental BJT structure on SOI to implement our solution. We will also demonstrate that the use of SiGe base and a Schottky collector in the proposed structure will further enhance the transistor performance making it a suitable candidate for high speed BiCMOS applications involving compatible NPN and PNP transistors with wide bandgap SiC-emitters. We also have suggested in the end a possible fabrication procedure for the device using well established fabrication steps for lateral BJTs. It may be pointed out that the actual fabrication will not be without its problems. However, we believe that the proposed structure may provide significant incentive for experimental exploration considering the importance of wide bandgap emitter PNP transistors in many circuit applications.

## 2. Comparison of $V_{CE(offset)}$ problem in wide bandgap emitter NPN/PNP HBTs

The collector-emitter offset voltage is defined as the difference between the turn-on voltage of the emitter-base (EB) and base-collector (BC) junctions [11]:

$$V_{CE(offset)} = V_{EB} - V_{BC} \text{ at } I_C = 0 \text{ mA}. \quad (1)$$

In order to compare the collector-emitter off-set problem in wide bandgap PNP/NPN HBTs, we have chosen a lateral HBT structure as shown in Fig. 1 in which the emitter region is SiC, base and collector regions are silicon. The epitaxial film thickness of the HBT is chosen to be 0.2 μm and the buried oxide thickness is 0.38 μm. Emitter is doped at $5\times10^{19}$ cm$^{-3}$. Base width is 0.4 μm and its doping is chosen to be $5\times10^{17}$ cm$^{-3}$. Collector is doped at $2\times10^{17}$ cm$^{-3}$. All these parameters are chosen based on reported experimental results of lateral silicon NPN BJTs [12]. The collector-emitter offset problem of SiC emitter PNP HBTs can be best understood from Fig. 2 in which the emitter-base (EB) and base-collector (BC) junction diode characteristics are shown. Fig. 2(a) shows that the turn on voltage $V_{EB}$ for emitter-base junction of the PNP SiC emitter HBT is 1.5 V larger than that of the emitter-base junction of the NPN SiC emitter HBT. On the other hand the turn on voltage $V_{BC}$ of the collector-base junction for both the SiC emitter PNP/NPN HBTs is identical (~ 0.8 V) as shown in Fig. 2(b) since silicon is used for both the base and collector regions. The output characteristics of the PNP and NPN wide bandgap SiC emitter HBTs are shown in Fig. 3. It can be clearly observed from Fig. 3(a) that the SiC emitter PNP HBT exhibits a large collector-emitter offset voltage ~(1.75 V) compared to the SiC emitter NPN HBT due to the large built-in voltage difference between the emitter-base (EB) and base-collector (BC) junctions [8-9, 11] seen in Fig. 2. According to eq.(1), the collector-emitter offset voltage for SiC emitter NPN HBT is expected to be of ~(0.25 V) and matches well with the offset voltage shown in Fig.3(b). The collector-emitter offset voltage of the SiC emitter PNP HBT calculated using eq.(1) comes out to be ~(1.75 V) and matches well with the offset voltage predicted from the IV characteristics shown in Fig. 3. Due to this large collector-emitter offset-voltage exhibited by the SiC emitter PNP HBTs, they cannot be used along with SiC emitter NPN HBTs. It can also be observed from Fig. 3 that the wide bandgap SiC emitter PNP HBT has a lesser current gain than that of the wide bandgap NPN HBT. But often, it is essential to have the performance of PNP HBTs nearly identical to that of NPN HBTs in BiCMOS applications such as push-pull amplifier design and also in ECL and complementary (npn/pnp) logic design.

## 3. Dual bandgap emitter approach to reduce $V_{CE}$(offset) in SiC emitter PNP HBTs

In this section, we show that the collector-emitter offset-voltage of wide bandgap SiC emitter PNP transistors can be significantly reduced by replacing the SiC emitter in the lateral PNP HBT by a dual bandgap emitter consisting of SiC on Si as shown in Fig. 4. The presence of a thin layer of P-silicon in the emitter reduces the collector-emitter offset-voltage drastically by eliminating built-in potential difference between emitter-base (EB) and base-collector (BC) junctions. The reduction of collector-emitter offset voltage in dual bandgap emitter PNP HBT can be best understood from Fig. 5. As can be observed from Fig. 5, there is a large reduction in built-in potential of emitter-base (EB) junction after replacing SiC emitter by the dual bandgap emitter. Now the theoretically predicted offset voltage for the dual bandgap emitter PNP HBT should be approximately ~(0.05 V) according eq.(1). The output characteristics of the dual bandgap PNP HBT (emitter thickness: 0.15 μm SiC + 0.05 μm Si) are compared in Fig. 6 with that of SiC wide bandgap P-emitter PNP HBT. We observe that the $V_{CE(offset)}$ of the dual bandgap emitter PNP HBT matches with the value calculated from eq.(1) and is significantly smaller than the $V_{CE(offset)}$ of the SiC P-emitter HBT. However, while the introduction of a thin layer of silicon in the wide bandgap emitter reduces $V_{CE(offset)}$, it is also accompanied by a reduction in the current gain. In Fig. 7, the dependence of current gain is shown for different relative values of SiC and silicon in the emitter region. We notice that as the si film thickness increases, the current gain decreases. We demonstrate in the following section that the loss in current gain can be recovered by introducing the SiGe base in the proposed structure shown in Fig. 4.

## 4. Application of SiGe Base to the dual bandgap emitter PNP HBT

To improve the current gain of the dual bandgap emitter PNP HBT, in our simulations we have replaced the silicon base by the SiGe base(20% Ge content) in the proposed structure. The simulated current gain of the dual bandgap emitter with and without the SiGe base is shown in Fig. 8. The presence of SiGe in the base region improves the emitter injection efficiency resulting in a higher current gain. Although the application of SiGe base restores the current gain, it is well known that PNP transistors suffer from large collector resistance due to low hole mobility. Further, SiGe base transistors suffer from the additional problem of excess stored base charge due to the accumulation of carriers at the collector-base junction [13]. The application of a Schottky collector has been shown to improve the performance of PNP bipolar transistors [8, 14-15]. Therefore, it would be of great interest to see how the usage of the Schottky collector to the dual bandgap emitter SiGe base PNP transistor will enhance its performance.

## 5. Application of Schottky collector to the dual bandgap emitter SiGe base PNP HBT and its impact on device performance

In order to study the effect of the Schottky collector, we have replaced the P-collector of the proposed structure shown in Fig. 4 with a Schottky contact. Based on experimental results, it has been reported [16] that platinum silicide gives the highest barrier height ($\phi_{Bn}$ = 0.82 eV) with n-SiGe base. Therefore, making an appropriate Schottky contact to the n-SiGe base is not a difficult task.

The Gummel plots of this dual bandgap SiC-on-Si P-emitter SiGe base HBT with and without the Schottky collector transistor are compared in Fig. 9. We observe that the base current in the dual bandgap emitter SiGe base lateral Schottky collector PNM HBT is smaller than that of the dual bandgap emitter SiGe base PNP HBT. This is mainly because of finite electron current $I_{nm}$ caused by the electron flow from metal into the n-base [17]. As the electron current from emitter to base is fixed by the emitter-base forward bias voltage, the electron current $I_{nm}$ from metal to n-base flows into the base terminal [17] reducing the total base current. As a result of this, the current gain of the dual bandgap emitter SiGe base lateral Schottky collector PNM HBT is higher than that of the dual bandgap emitter SiGe base PNP HBT as shown in Fig. 10. An interesting point is that the base current of the dual bandgap emitter SiGe base lateral Schottky collector PNM HBT is less than that of the dual bandgap emitter SiGe base PNP HBT even at high–level injection of carriers as shown in Fig. 9 which clearly shows the suppression of the Kirk effect [18] in the dual bandgap emitter SiGe base lateral Schottky collector PNM HBT. The simulated I-V characteristics of dual bandgap emitter SiGe base lateral Schottky collector PNM HBT and dual bandgap emitter SiGe base PNP HBT are shown in Fig. 11. As can be seen, the current-voltage characteristics of the dual bandgap emitter SiGe base lateral Schottky collector PNM HBT are superior to those of the dual bandgap emitter SiGe base PNP HBT in terms of reduced collector resistance. However, there is a finite offset voltage for the dual bandgap emitter SiGe base Schottky collector PNM HBT mainly due to the reduced built-in potential of base-collector Schottky junction.

Fig. 12 shows the transient behaviour of the dual bandgap emitter SiGe base lateral Schottky collector PNM HBT compared with the dual bandgap emitter SiGe base PNP HBT. It is clear that the dual bandgap emitter SiGe base lateral Schottky collector PNM HBT exhibits excellent transient response with nearly zero base charge storage time due to its metal collector and suppressed Kirk effect while compared dual bandgap emitter SiGe base PNP HBT shows a higher storage time due to the Kirk effect and also the electron pile-up at the collector-base hetero-junction [13, 18].

Fig. 13 shows the unity gain cut-off frequency versus collector current of the dual bandgap emitter SiGe base lateral Schottky collector PNM HBT and is compared with the dual bandgap emitter SiGe base PNP HBT. As can be observed, the cut-off frequency of the dual bandgap emitter SiGe base lateral Schottky collector PNM HBT is higher than that of dual bandgap emitter SiGe base PNP HBT due to its metal collector and higher transconductance $g_m$. The dual bandgap emitter SiGe base lateral Schottky collector PNM HBT exhibits an $f_T$ of 3.55 GHz at a collector current of 0.6 mA, whereas for the comparable dual bandgap emitter SiGe base PNP HBT, $f_T$ falls to a negligible value at the above current due to Kirk effect and decrease in transconductance.

## 6. The effect of doping and Ge % in the base

In all the above simulations we have assumed the base Ge concentration to be 20% which is the practical upper limit on Ge in most practical applications. However, it will be interesting to see the how the current gain and the breakdown voltage of the proposed structure change if the Ge concentration in the base region is varied. If ion-implantation is used to create the SiGe base, it is quite possible that the Ge content may lie in the range of 10 to 12 %. Therefore, we have next investigated the effect of base doping on peak current gain and breakdown voltage $BV_{CEO}$ (for zero base current) for various germanium concentrations in the SiGe-base of the dual bandgap emitter SiGe base lateral Schottky collector PNM HBT.

Fig. 14 shows the peak current gain versus base doping for various germanium concentrations in the SiGe-base of the dual bandgap emitter SiGe base lateral Schottky collector PNM HBT. It can be observed from this figure that the peak current gain decreases as we decrease the germanium concentration in the SiGe base for a given base doping and the gain also decreases as we increase the base doping for a given Ge concentration because of low emitter injection efficiency. Fig. 15 shows the breakdown voltage $BV_{CEO}$ (for zero base current) versus base doping for various germanium concentrations in the SiGe base of the dual bandgap emitter SiGe base lateral Schottky collector PNM HBT. We note that for a given base doping, the breakdown voltage $BV_{CEO}$ (for zero base current) increases as we decrease the germanium concentration and for a given Ge concentration, the breakdown voltage increases as we increase the base doping due to increasing critical electric field. The above design curves provide useful indicators on the required Ge concentration and base doping to realize a given current gain and breakdown voltage.

## 7. Proposed fabrication procedure

Fabrication of the proposed structure can be realized by introducing a few extra steps in the reported fabrication procedure of the lateral BJTs on SOI [12]. We can start with an SOI wafer having n-type epitaxial layer thickness of 0.2 μm and doping of 5 x $10^{17}$ cm$^{-3}$. In the first step, a thick CVD oxide is deposited and patterned as shown in Fig. 16(a). The uncovered N-region is converted into P-region by implanting a p-type dopant at a calibrated tilt angle as discussed in [15] as shown in Fig. 16(b). The P-type emitter region is then etched to a thickness of 0.05 μm as shown in Fig. 16(c). In the next step, we deposit the $p^+$ SiC on the horizontal edge (at point X in Fig. 16(d)) of the silicon surface which acts as a seed and the SiC grows [19-20] as shown in Fig. 16(d). Subsequent to this step, CMP process is performed and then a thick CVD oxide is deposited and patterned as shown in Fig. 16(e). Following this step, a nitride film is deposited as shown in Fig. 16(f). In the next step, an unmasked RIE etch is performed until the planar silicon nitride is etched. This retains the nitride spacer at the vertical edge of thick CVD oxide as shown in Fig. 16(g). After a thick oxide is deposited as shown in Fig. 16(h), CMP process is carried out to planarize the surface. Next, nitride spacer is removed with selective etching, which will create a window in the oxide as shown in Fig. 16(i). Germanium can now be implanted [21-24] through this window to convert silicon in the base region to SiGe. Ge implantation can be performed at an energy of 130 KeV with fluences of 1, 2, or 3 x $10^{16}$cm$^{-2}$ according to the reported works in the literature [21]. To re-crystallize the implanted SiGe layer, a rapid thermal annealing (RTA) need to be performed at 1000 $^o$C for about 10 s. This process is to ensure complete re-crystallization of SiGe amorphous layer [21].

After converting, silicon in the base region to SiGe, we then deposit $n^+$ - poly and then the wafer is once again planarized using CMP leaving $n^+$ - poly in the place where the nitride film was present earlier as shown in Fig. 16(j). Following this step, a contact window is opened for metal Schottky collector as shown in Fig. 16(k) and subsequent to this step, the $p^+$ emitter contact window is opened as shown in Fig. 16(l). Finally, platinum silicide is deposited to form the Schottky collector contact and ohmic contacts on the emitter and $p^+$ - poly base region.

## 8. Conclusion

In this paper, we have first discussed the reasons for the significant collector-emitter offset voltage observed in wide bandgap SiC P-emitter HBTs. Based on numerical simulations, we have demonstrated that using a dual bandgap SiC-on-Si emitter in the presence of the SiC P-emitter, greatly reduces the collector-emitter offset voltage of wide bandgap PNP HBTs. However, the presence of Si in the emitter results in a reduced current gain and the low hole mobility in the P-collector gives rise to a high collector resistance. To overcome this problem, we have applied the SiGe base and a metal Schottky collector to the proposed structure and demonstrated that the resulting device not only will have very low collector emitter offset voltage but will also exhibit high current gain and negligible storage time. Based on reported experimental results for the lateral BJTs on SOI, we have also suggested a possible fabrication procedure for the proposed structure. We conclude from our study of the dual bandgap SiC P-emitter HBT with the combination of SiGe base and Schottky collector that the proposed structure should be a good candidate for BiCMOS applications requiring both NPN and PNP HBTs with comparable performance.

## Acknowledgement

Financial support from Department of Science and Technology (DST), Government of India is gratefully acknowledged. We also would like to thank Council of Scientific and Industrial Research (CSIR), Government of India, for the fellowship given to Mr. Linga Reddy.

# Figure captions

Fig. 1. Cross-sectional view of the wide bandgap SiC emitter lateral PNP/NPN HBT.

Fig. 2. (a). E-B and (b). B-C diode characteristics of the SiC emitter NPN HBT and SiC emitter PNP HBT.

Fig. 3. Common-emitter IV – characteristics of (a). SiC emitter PNP HBT and (b). SiC emitter NPN HBT.

Fig. 4. Cross-sectional view of the dual bandgap emitter lateral PNP HBT.

Fig. 5. E-B and B-C diode characteristics of the dual bandgap emitter PNP HBT.

Fig. 6. Common-emitter IV – characteristics of the SiC emitter PNP HBT compared with that of the dual bandgap emitter PNP HBT.

Fig. 7. Gain versus collector current characteristics of the dual bandgap emitter PNP HBT for various thicknesses of emitter ($V_{CE}$ = -1 V).

Fig. 8. Gain versus collector current characteristics of the dual bandgap emitter PNP HBT with and without SiGe base ($V_{CE}$ = -1 V).

Fig. 9. Gummel plot of the dual bandgap emitter SiGe base lateral Schottky collector PNM HBT compared with that of the dual bandgap emitter SiGe base PNP HBT ($V_{CE}$ = -1 V).

Fig. 10. Gain versus collector current characteristics of the dual bandgap emitter SiGe base lateral Schottky collector PNM HBT compared with that of the dual bandgap emitter SiGe base PNP HBT ($V_{CE}$ = -1 V).

Fig. 11. Common-emitter IV – characteristics of the dual bandgap emitter SiGe base lateral Schottky collector PNM HBT compared with that of the dual bandgap emitter SiGe base PNP HBT.

Fig. 12. Transient behaviour of the dual bandgap emitter SiGe base lateral Schottky collector PNM HBT compared with that of the dual bandgap emitter SiGe base PNP HBT.

Fig. 13. Unity gain cut-off frequency versus collector current of the dual bandgap emitter SiGe base lateral Schottky collector PNM HBT compared with that of the dual bandgap emitter SiGe base PNP HBT.

Fig. 14. Gain versus base doping for various germanium concentrations in the base of the dual bandgap emitter SiGe base lateral Schottky collector PNM HBT.

Fig. 15. Breakdown voltage $BV_{CEO}$ (for zero base current) versus base doping for various germanium concentrations in the base of the dual bandgap emitter SiGe base lateral Schottky collector PNM HBT.

Fig. 16. Proposed fabrication steps for the dual bandgap emitter SiGe base lateral Schottky collector PNM HBT on SOI.

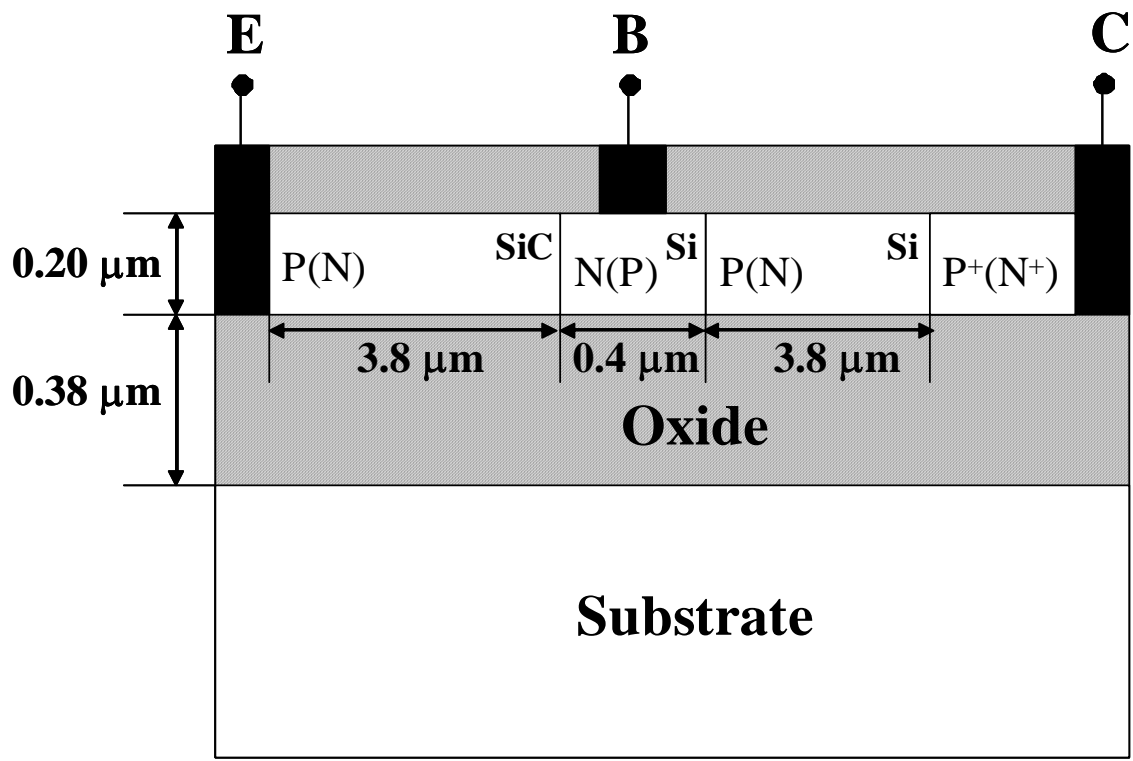

Fig. 1

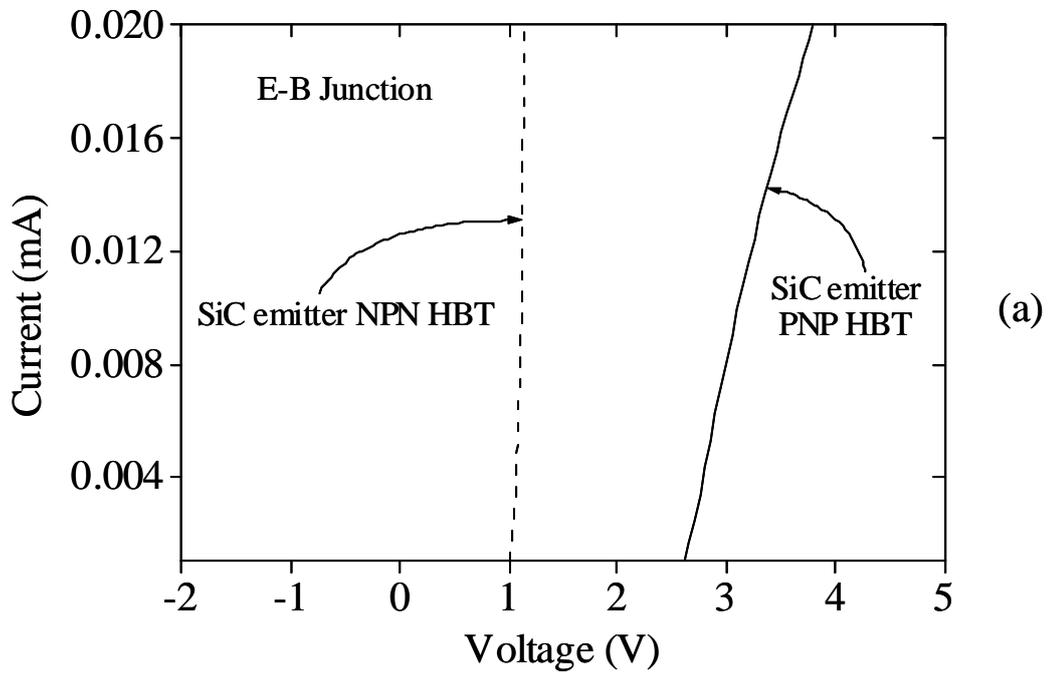

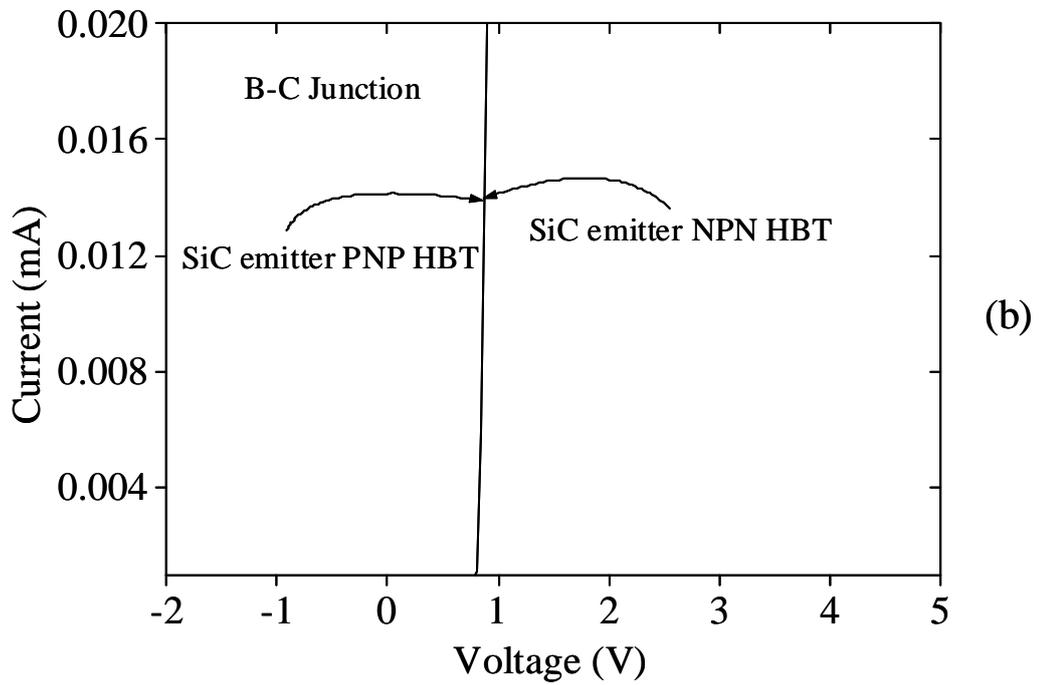

Fig. 2

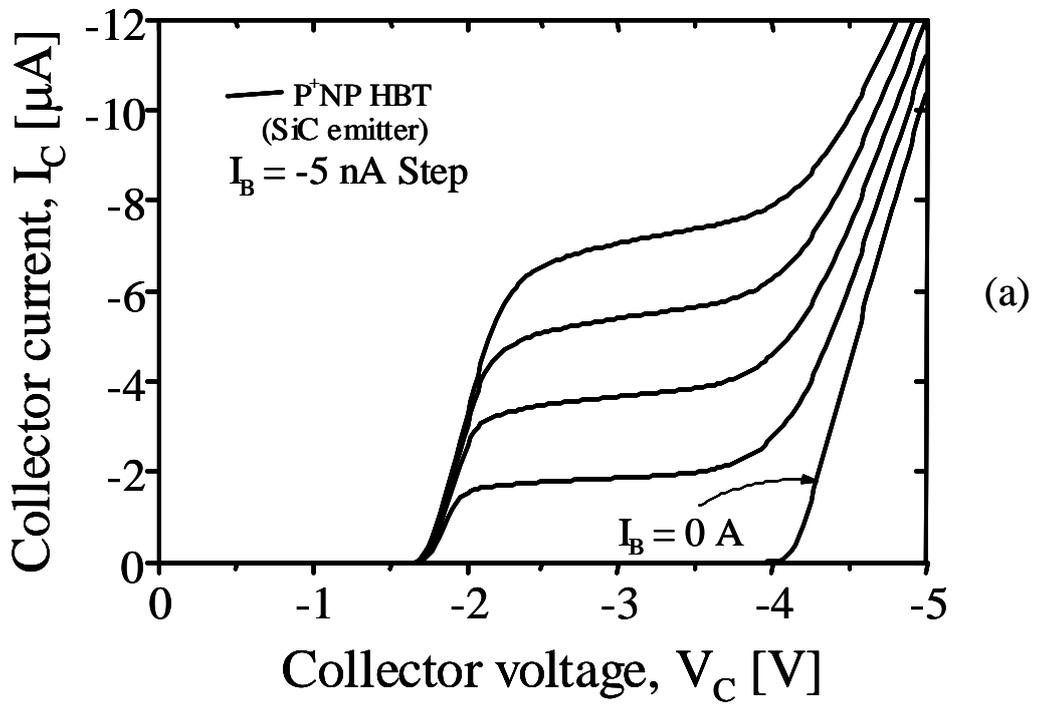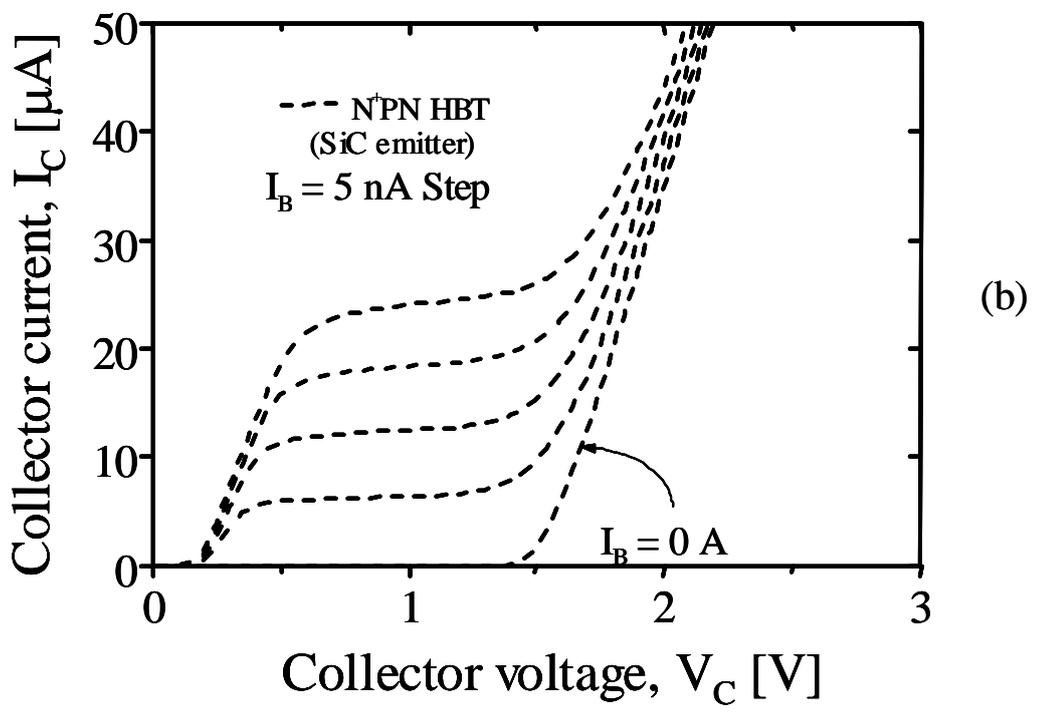

Fig. 3

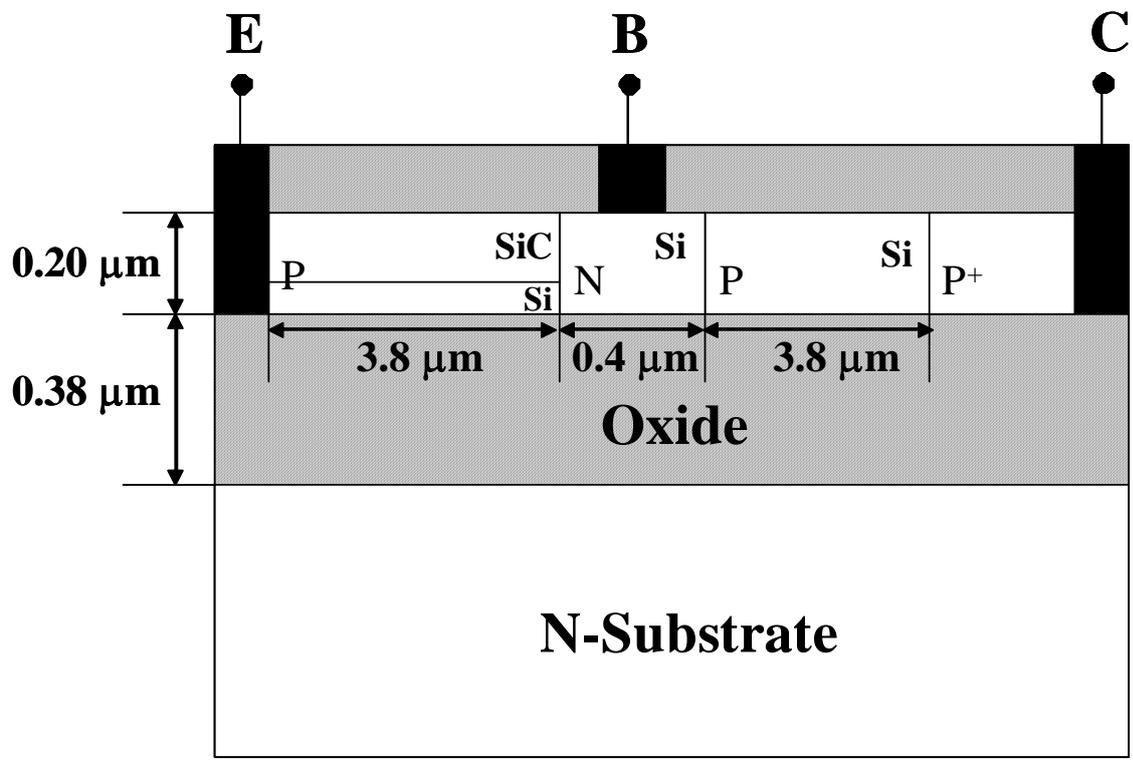

Fig. 4

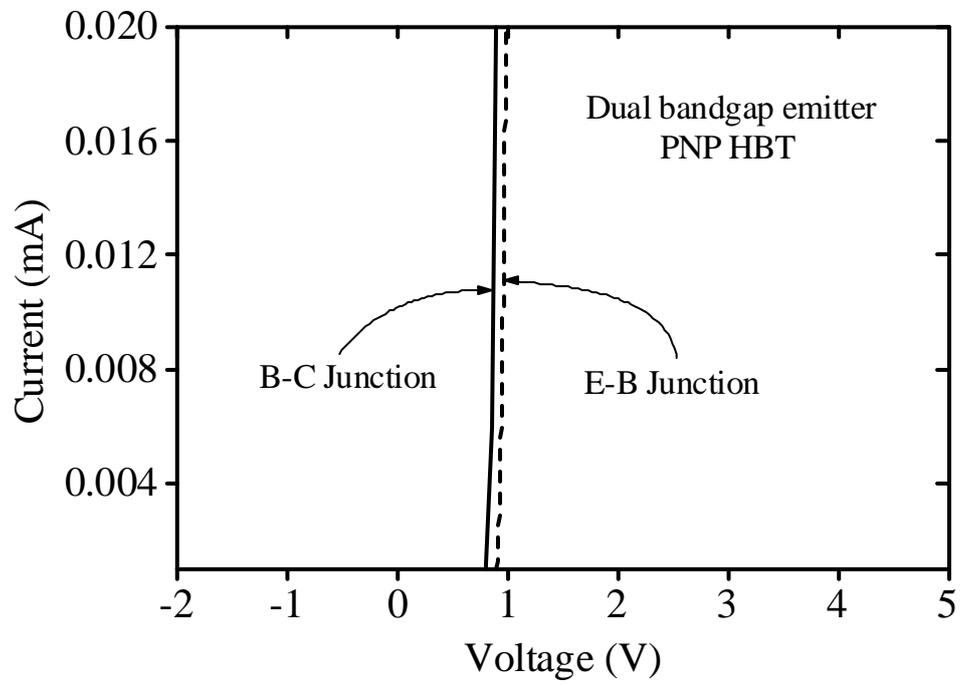

Fig. 5

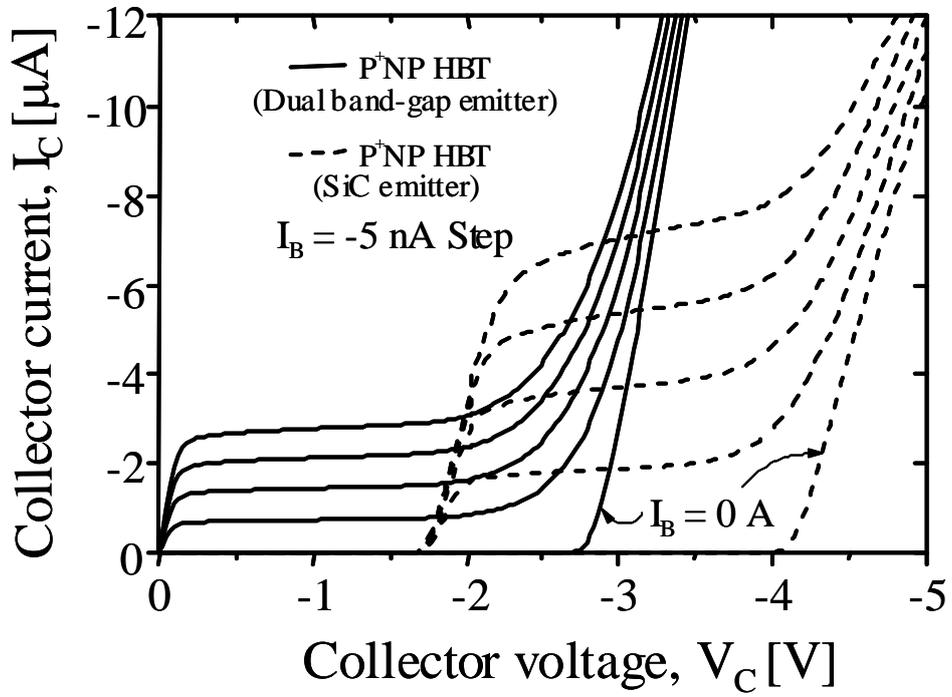

Fig. 6

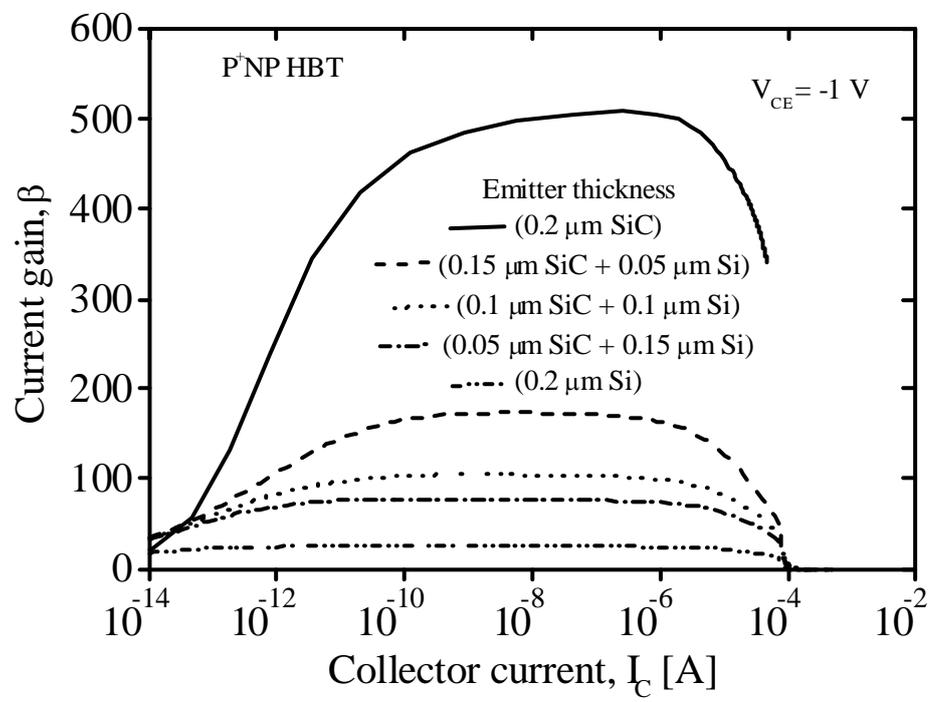

Fig. 7

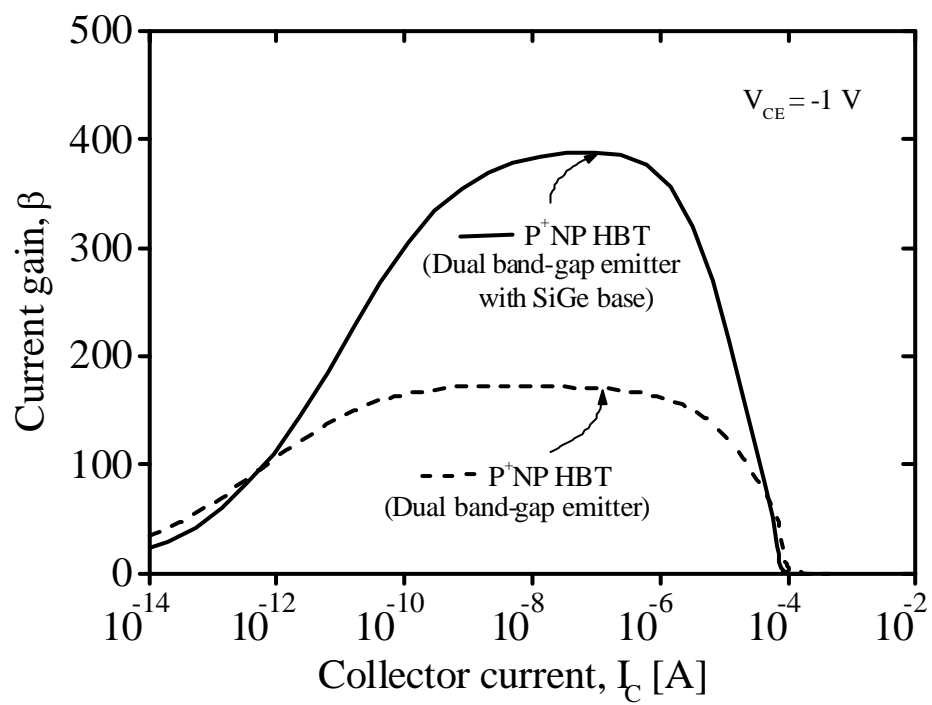

Fig. 8

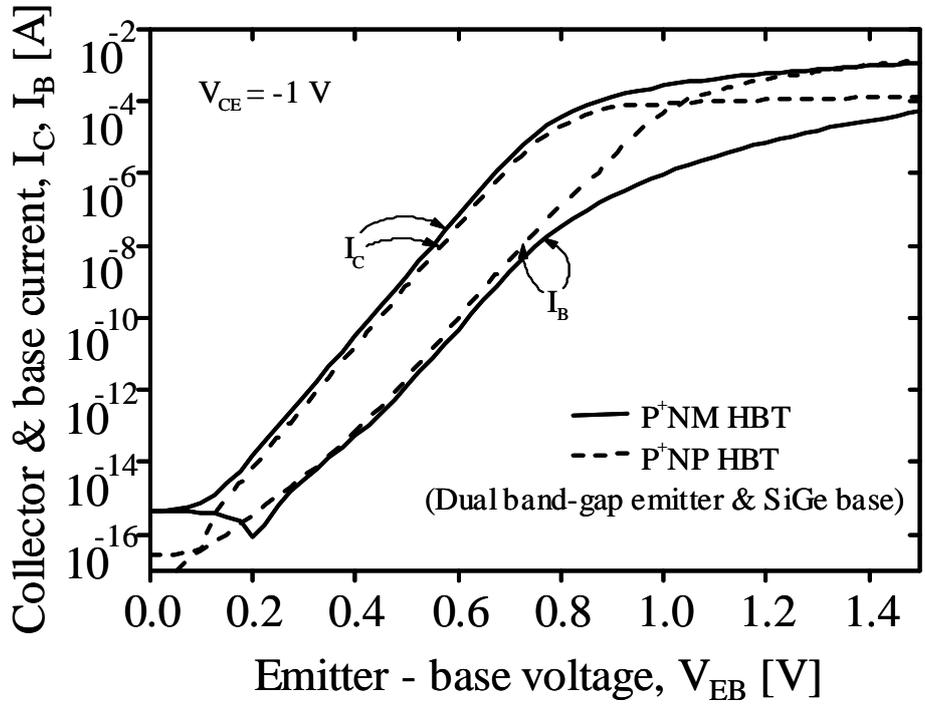

Fig. 9

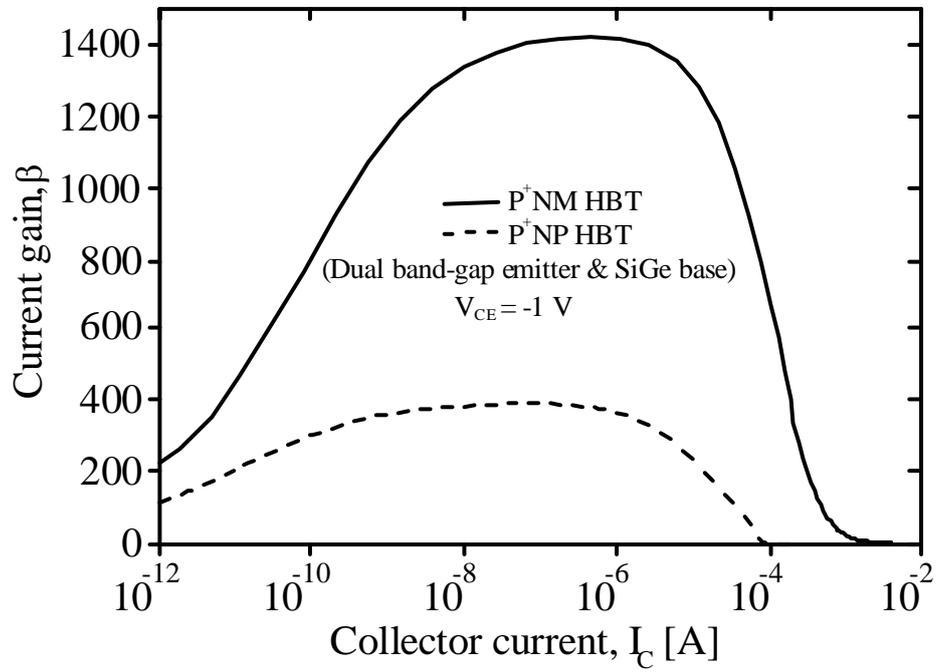

Fig. 10

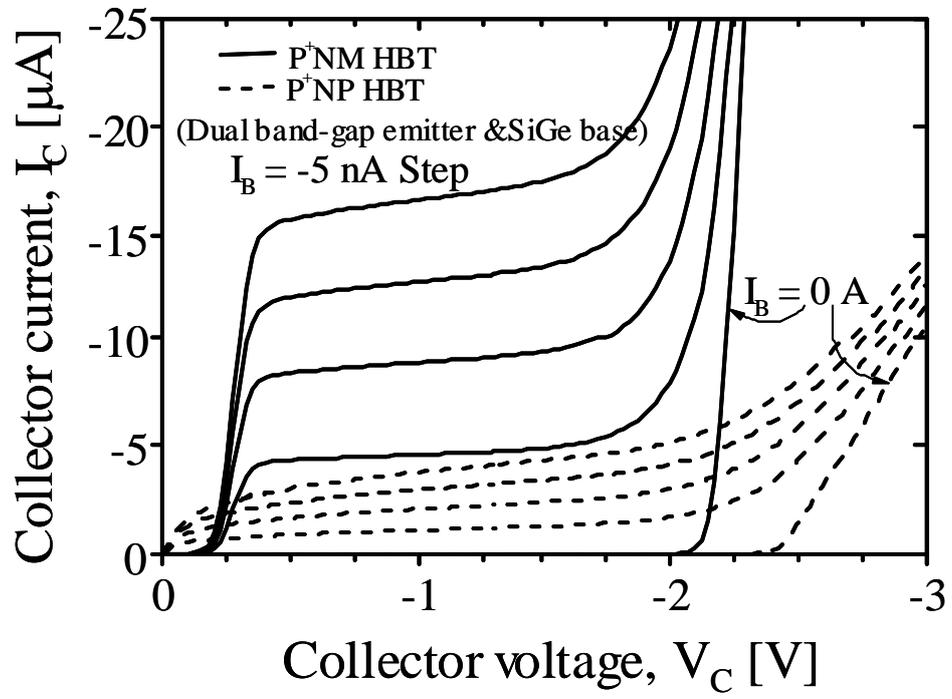

Fig. 11

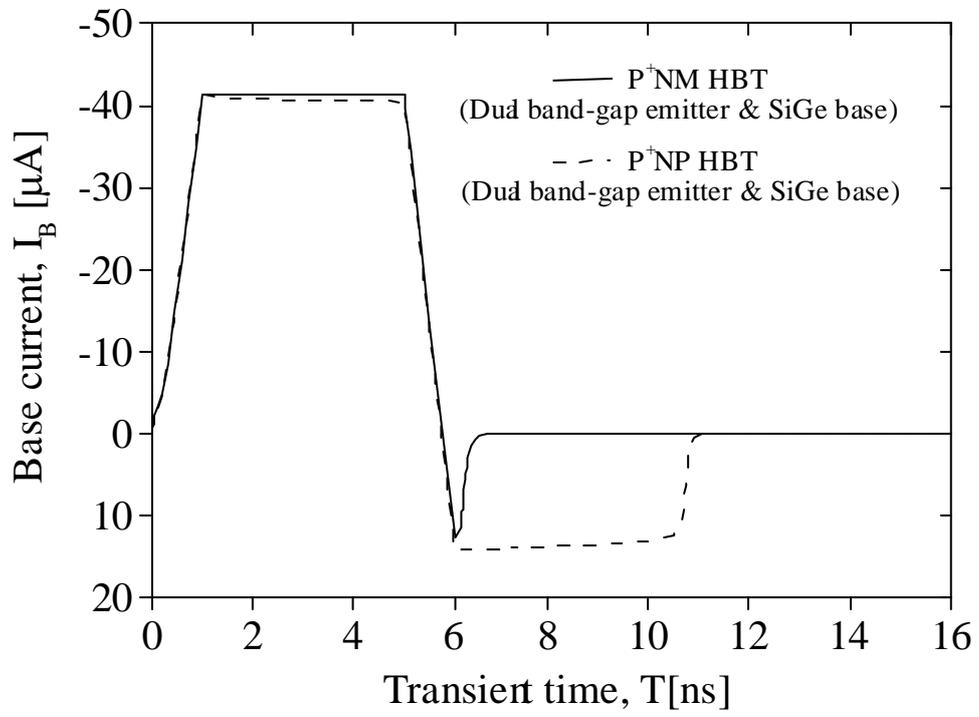

Fig. 12

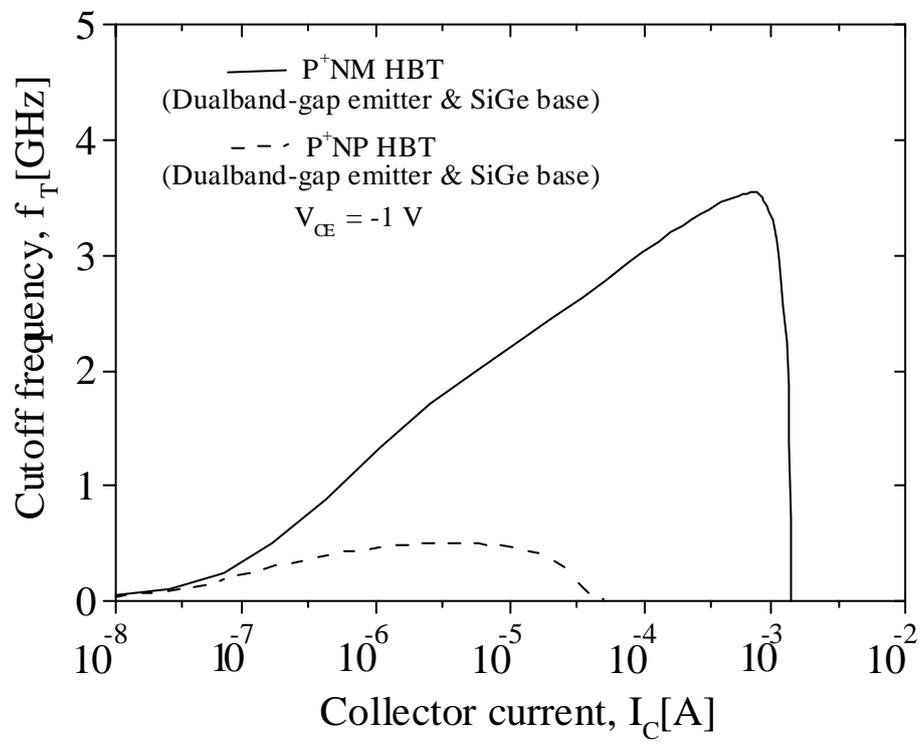

Fig. 13

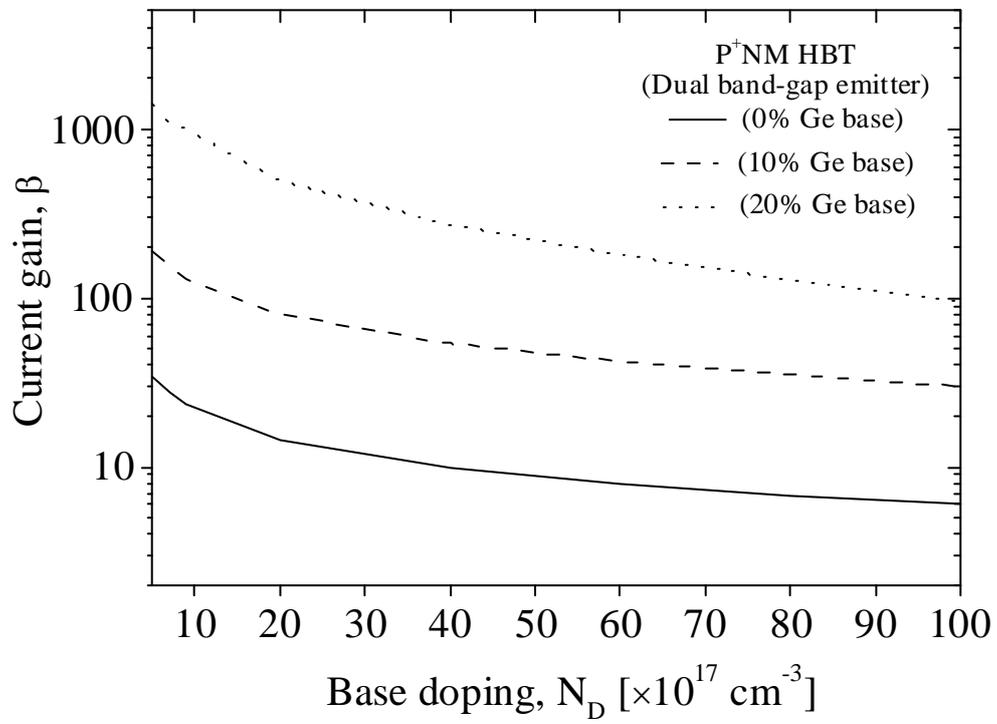

Fig. 14

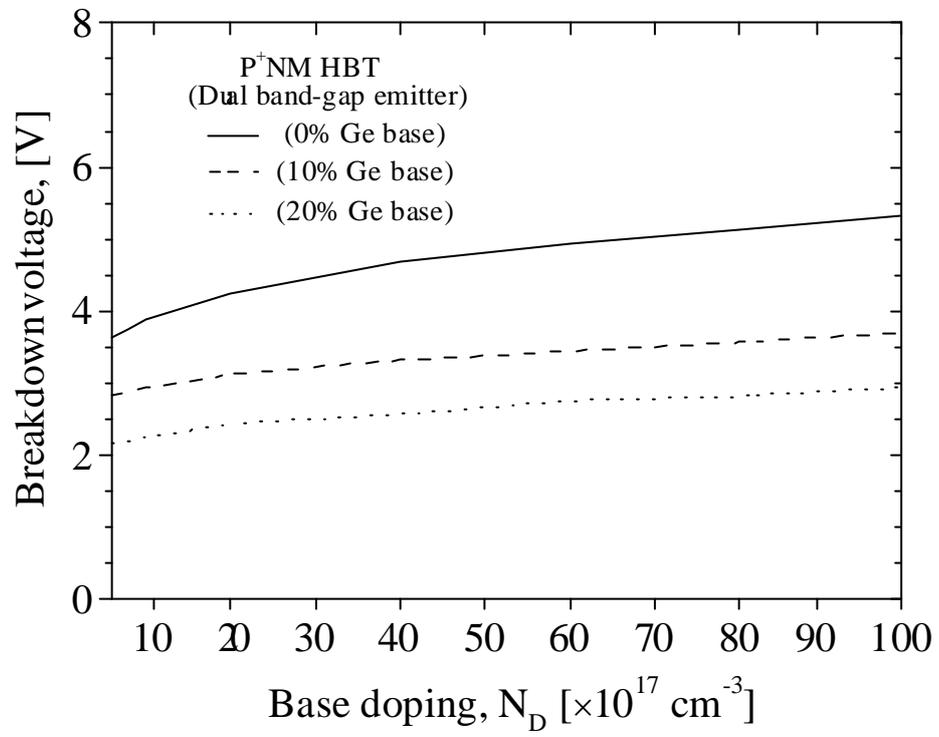

Fig. 15

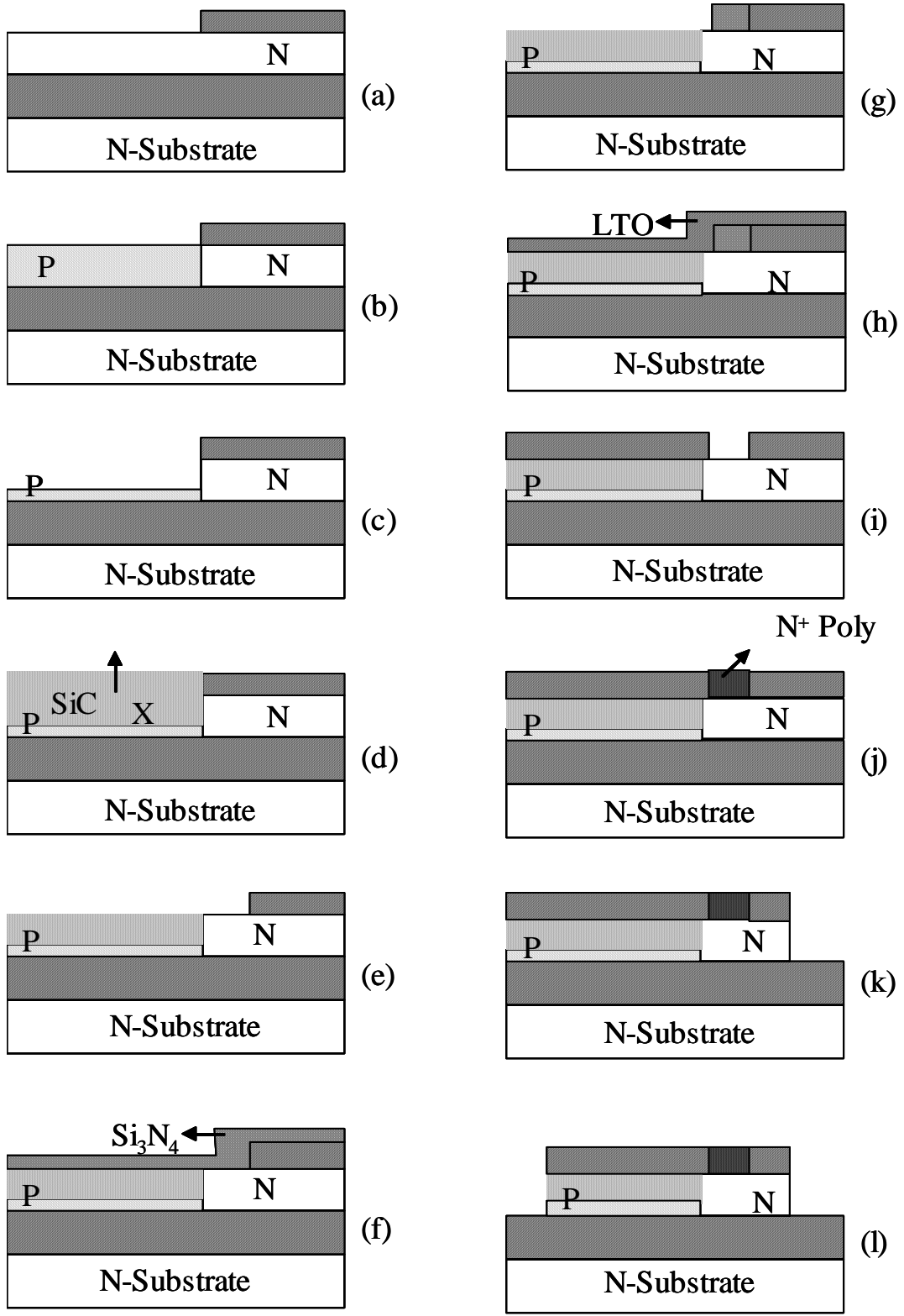

Fig. 16